%% file: Herrnstein.tex
\documentclass{aastex}
\usepackage{emulateapj5}

\usepackage{natbib}

\newcommand{\bq}{\begin{equation}}
\newcommand{\eq}{\end{equation}}

\newcommand{\p}{$\pm$}

\newcommand{\simgt}{\lower.5ex\hbox{$\; \buildrel > \over \sim \;$}}
\newcommand{\simlt}{\lower.5ex\hbox{$\; \buildrel < \over \sim \;$}}
\usepackage{natbib}

\begin{document}
 
\title{The Variability of Sagittarius~A* at Centimeter Wavelengths}
 
\author{
Robeson M. Herrnstein\altaffilmark{1},
Jun-Hui Zhao\altaffilmark{2}, 
Geoffrey C. Bower\altaffilmark{3}, and
W. M. Goss\altaffilmark{4}}

\altaffiltext{1}{Department of Astronomy, Columbia University, Mail Code 5246, 550 West 120th St., New York, NY 10027, herrnstein@astro.columbia.edu}
\altaffiltext{2}{Harvard-Smithsonian Center for Astrophysics, 60 Garden Street, Cambridge, MA 02138, jzhao@cfa.harvard.edu}
\altaffiltext{3}{601 Campbell Hall, Radio Astronomy Lab, UC Berkeley, Berkeley, CA 94720, gbower@astron.berkeley.edu}
\altaffiltext{4}{National Radio Astronomy Observatory, P.O. Box 0, Socorro, NM, 87801, mgoss@aoc.nrao.edu }

\slugcomment{Accepted for publication in AJ}

\begin{abstract}
We present the results of a 3.3-year project to monitor the flux
density of Sagittarius~A* at 2.0, 1.3, and 0.7~cm with the Very Large
Array.  Between 2000.5 and 2003.0, 119 epochs of data were taken with
a mean separation between epochs of eight days.  After 2003.0,
observations were made roughly once per month for a total of nine
additional epochs.  Details of the data calibration process are
discussed, including corrections for opacity and elevation effects as
well as changes in the flux density scales between epochs.  The fully
calibrated light curves for Sgr~A* at all three wavelengths are
presented.  Typical errors in the flux density are 6.1\%, 6.2\%, and
9.2\% at 2.0, 1.3, and 0.7~cm, respectively.  There is preliminary
evidence for a bimodal distribution of flux densities, which may
indicate the existence of two distinct states of accretion onto the
supermassive black hole.  At 1.3 and 0.7~cm, there is a tail in the
distribution towards high flux densities.  Significant variability is
detected at all three wavelengths, with the largest amplitude
variations occurring at 0.7~cm.  The rms deviation of the flux density
of Sgr~A* is 0.13, 0.16, and 0.21~Jy at 2.0, 1.3, and 0.7~cm,
respectively.  During much of this monitoring campaign, Sgr~A*
appeared to be relatively quiescent compared to results from previous
campaigns.  At no point during the monitoring campaign did the flux
density of Sgr A* more than double its mean value.  The mean spectral
index of Sgr~A* is $\alpha=0.20\pm0.01$ (where
$S_\nu\propto\nu^\alpha$), with a standard deviation of 0.14.  The
spectral index appears to depend linearly on the observed flux density
at 0.7~cm with a steeper index observed during outbursts.  This
correlation is consistent with the expectation for outbursts that are
self-absorbed at wavelengths of 0.7~cm or longer and inconsistent with
the effects of simple models for interstellar scintillation.  Much of
the variability of Sgr~A*, including possible time lags between flux
density changes at the different wavelengths, appears to occur on time
scales less than the time resolution of our observations (8 days).
Future observations should focus on the evolution of the flux density
on these time scales.

\end{abstract}

\keywords{Galaxy: center --- radio continuum: galaxies --- accretion --- black hole physics}

\section{Introduction \label{intro}}

Observations of stellar proper motions in the central $1''$ (0.04~pc)
of the Milky Way suggest that a $4\times10^6$~M$_\odot$ black hole is
located at the dynamical center of the Galaxy \citep{sch02,ghe03b}.
In the radio, the bright ($\sim1$~Jy), compact source called
Sagittarius A* (Sgr~A*) appears to be closely associated with the
supermassive black hole \citep{men97}.  For over two decades, the
radio flux density of Sgr~A* has been known to vary \citep{bro82}, but
the cause of this variability remains unclear.  The radio variability
tends to increase towards shorter wavelengths, and significant fluctuations
on roughly weekly time scales are observed at wavelengths of 2~cm and
shorter \citep{zha01}.  Based on this observed increase in fractional
variability towards shorter wavelengths, the variability of Sgr~A* at
centimeter wavelengths has been suggested to be intrinsic to the
source \citep{zha92}. At millimeter and sub-millimeter wavelengths,
the flux density of Sgr~A* is even more variable \citep{wri93,
tsu99,zha03}.  Changes of up to a factor of four in the flux density
have been observed at 1.3~mm using the Sub-Millimeter Array (SMA)
\citep{zha03}.

Short-term variability of Sgr~A* has been detected in X-rays.  In
2002, a multi-wavelength campaign was undertaken to simultaneously
monitor Sgr~A* at centimeter, millimeter, infra-red, and X-ray
wavelengths.  Based on 500 kilo-seconds of Chandra observations,
\citet{bag03} finds that flares of a factor of $\sim10$ occur roughly
once per day.  (In this paper, we call events on hour-to-day time
scales ``flares'', while events on time scales of weeks are referred
to as ``outbursts''.)  The strongest X-ray flare during this
observation occurred on 29 May 2002 and showed a factor of 20 increase
in the flux of Sgr~A*.  To date, three additional large X-ray flares
have been observed between June 2000 and October 2003.  On 27 October
2000, the flux of Sgr~A* detected by Chandra increased by a factor of
45 during a 10~ksec flare \citep{bag01}.  A factor of 20 increase in
the flux of Sgr~A* at 2--10~keV was detected by XMM-Newton during the
beginning of a flare on 4 September 2001, but the observations
unfortunately did not include the entire event \citep{goldwurm03}.
More recently, a factor of 160 flare lasting 2.7~ks was detected with
XMM-Newton on 3 October 2002 \citep{porquet03}.

Recently, Sgr~A* has been detected for the first time at infrared
wavelengths.  In the infrared, Sgr~A* appears to show both long-term
and short-term variability.  The short-term variability occurs on time
scales of $\sim1$~hr, similar to the time scale of X-ray events
\citep{gen03}.  The day to week time scale of the long-term infrared
variability is more similar to the time scale of radio outbursts
\citep{ghe03b}.

In order to understand the detailed nature of the variability of
Sgr~A* in the radio, \citet{zha01} (hereafter ZBG01) combined Very
Large Array\footnote{The National Radio Astronomy Observatory is a
facility of the National Science Foundation operated under cooperative
agreement by Associated Universities, Inc.}  (VLA) data from frequent
monitoring of Sgr~A* between 1990.1 and 1993.5 \citep{zha92} with
twenty years of additional archival data from 1977 to 1999.  At
1.3~cm, the flux density of Sgr~A* typically varied by 30\%, and
occasional outbursts of 100\% were observed.  A comparison of the
light curves at 20, 6.0, 3.6, 2.0, and 1.3~cm showed that the largest
amplitude variations in flux density occurred at the shortest
wavelengths (ZBG01).

Using a Maximum Entropy Method \citep{pre89} as well as a classic
periodogram augmented with CLEAN to search for periodicities, ZBG01
found a periodicity of $106\pm10$ days ($1.1(\pm0.1)\times10^{-7}$~Hz)
at 3.6, 2.0, and 1.3~cm.  The light curves were consistent with no
phase offset between the three wavelengths.  The mean profile of the
106 day cycle at 1.3~cm had a broad peak, roughly 25\p5 days wide at
full width at half maximum (FWHM), indicating that the variability of
Sgr~A* is most likely quasi-periodic in nature.  Analysis of data
presented in ZBG01 was hindered, however, by highly irregular sampling
intervals.  Observations were only made on a regular basis (with
sampling intervals from 1 to 28 days) between 1990.1 and 1991.5, a
small fraction of the entire study.  Between 1991.5 and 1993.5, the
sampling was less frequent with a maximum sampling interval of 120
days.  Gaps in the archival data were as large as 1200 days.

Additional evidence for periodic variability on the order of 100 days
has been detected at 2 and 3~mm using 46 epochs of data taken between
1996 and 2000 with the Nobeyama Millimeter Array (NMA) \citep{tsu02}.
When the data are folded with a period of 106 days, the resulting
cycle divides roughly equally into a ``high'' and ``low'' activity
state.  During the high activity state, the flux density of Sgr~A* can
vary by as much as a factor of three, while Sgr~A* is relatively
quiescent in the low activity state.

Evidence for a period of 57 days was suggested by \citet{fal99} based
on 11~cm data from the Green Bank Interferometer (GBI).  This result
suggested that the 106-day periodicity could be a harmonic of a higher
frequency periodicity.  Due to sparse sampling over much of the 20~yr
baseline, the data presented by ZBG01 were not sensitive to periods
shorter than $\sim50$ days.  A new monitoring campaign in which the
flux density of Sgr~A* was regularly monitored on weekly time scales
was necessary to search for these short periodicities and detect the
shape of outbursts and phase offsets between the different wavelengths.

From June 2000 to October 2003, over 170 hours of VLA observing time
were used to monitor the flux density of Sgr~A* at 2.0, 1.3, and
0.7~cm.  The results of this 3.3-year project are presented in this
paper.  Section \ref{cal} explains in detail the data calibration,
including corrections for opacity effects and errors in absolute flux
calibration.  The fully calibrated flux densities for Sgr~A* at 2.0,
1.3, and 0.7 cm are presented in \S\ref{res} and are also listed in
their entirety in Table \ref{table}.  The characteristics of the light
curve, including the spectral index and time delay between
wavelengths, are also discussed.

This paper deals primarily with data calibration and the general
characteristics of the light curve and spectrum.  Results from the
analysis of these data for periodic or quasi-periodic signals will be
presented in subsequent papers.  Although \S \ref{res} briefly
compares our light curves to those from monitoring campaigns at other
wavelengths, detailed comparisons are presented elsewhere.
\citet{zha03} compares the radio light curve to preliminary data from
the Sub-Millimeter Array.  A separate paper also discusses evidence
for correlated events in the radio and X-ray and possible implications
for accretion models for Sgr~A* \citep{zha04}.  Finally, six Very Long
Baseline Array (VLBA) observations were made as part of this
monitoring campaign.  The results from these observations are
presented in \citet{bower04}.

\section{Observations}

Data were collected in the A, B, C, D, and hybrid configurations of
the VLA through projects AZ128 (21 June 2000 -- 26 September 2000),
AZ129 (5 October 2000 -- 27 September 2001), AZ136 (02 October 2001 --
04 January 2003), and AZ143 (13 January 2003 -- 14 October 2003).
From 2000.5 to 2003.0, the flux density of Sgr~A* at 2.0, 1.3, and
0.7~cm was measured roughly once per week (for a total of 119 epochs).
Between 2003.0 and 2003.8, nine additional observations were made once
per month in order to increase our sensitivity to long-term
periodicities.  Although a total of 128 observations were made,
problems with weather or instrumentation resulted in the occasional
loss of data at one or more wavelengths.  Successful flux densities
were measured for a total of 115 epochs at 2.0~cm, 124 epochs at
1.3~cm, and 121 epochs at 0.7~cm.  Sampling intervals in the finely
sampled data range from 1 to 26 days with a mean separation of eight
days (and a median separation of seven days).  Observations after
2003.0 have a mean and median separation of 30 days.  The resulting
data are sensitive to periods between roughly 15 and 2000 days.

A typical observation lasted a total of 1 to 2~hours.  Either 3C286 or
3C48 served as the primary flux calibrator and 1741--312 (J1744--312)
and 1817--254 (J1820--254) were used to track the phase during the
observations.  (B1950 name conventions are used throughout this
paper.)  Following an observation of the flux calibrator at 3.6~cm to
calibrate offsets in the pointing solutions, four minute integrations
were made at 2.0, 1.3, and 0.7~cm.  The pointing offsets were then
recalculated for 1741--312.  Because of the close proximity of the
sources, these solutions were also applied to Sgr~A* and 1817--254.
Observations of Sgr~A* at all three wavelengths were interleaved
between observations of a phase calibrator at roughly 15~minute
intervals.  A total of 5--15 minutes of integration time on Sgr~A*
were obtained at each wavelength for each observation.

\section{Data Reduction \label{cal}}
Initial calibration of the data was performed using the Astronomical
Image Processing System ({\it AIPS}).  Data from antennas and baselines
with large rms noise were flagged.  At all three wavelengths, the data
were corrected for changes in efficiency as a function of elevation
using the most recent gain curves for each antenna.  Prior to 25 July
2001, the gain curves at 2.0 and 1.3~cm were applied using the NRAO
supplied tasks, FIXUGAIN and FIXKGAIN.  At 0.7~cm, gain corrections
were applied using an algorithm we developed specifically for this
project.  The task also allowed us to correct the 0.7~cm data for
losses due to high opacity at low elevation. Because opacity effects
should be negligible at longer wavelengths, this correction was not
applied to the 2.0 and 1.3~cm data. Beginning 25 July 2001, both the
opacity and gain corrections were incorporated directly into the
{\it AIPS} task FILLM.  These corrections are equivalent to our previous
method and we subsequently used FILLM to apply opacity and gain
corrections to all of our data.

Primary flux density calibration was performed using 3C286 (1328+307),
except for 11 epochs where observations later than 20:00 LST forced us
to use 3C48 (0134+329).  When 3C48 was observed, baselines from
0--40~k$\lambda$ were used in the flux density calibration for all
three wavelengths.  For all other epochs, the flux density calibration
was performed using 3C286 and included all baselines longer than
150~k$\lambda$ at 2.0~cm or $185$~k$\lambda$ at 1.3~cm.  At both
wavelengths, solutions were calculated using 30~second integration
times.  At 0.7~cm, a model image of 3C286 obtained from C. Chandler of
NRAO was used in the flux density calibration, thus allowing the
inclusion of all baselines.  First, a phase-only flux density
calibration was performed using a solution interval equal to the
integration time (10 seconds).  The final calibration included both
amplitude and phase with a 30~second solution interval.  The flux
density scale calculated from 3C286 or 3C48 was applied to 1741--312
and 1817--254.  Phase calibrator flux densities and associated errors
for the phase calibrators were measured using the {\it AIPS} task
GETJY.

\begin{figure}
%\epsscale{0.8}
%\plotone{Herrnstein.fig1.eps}
\centering
\includegraphics[width=4.5in]{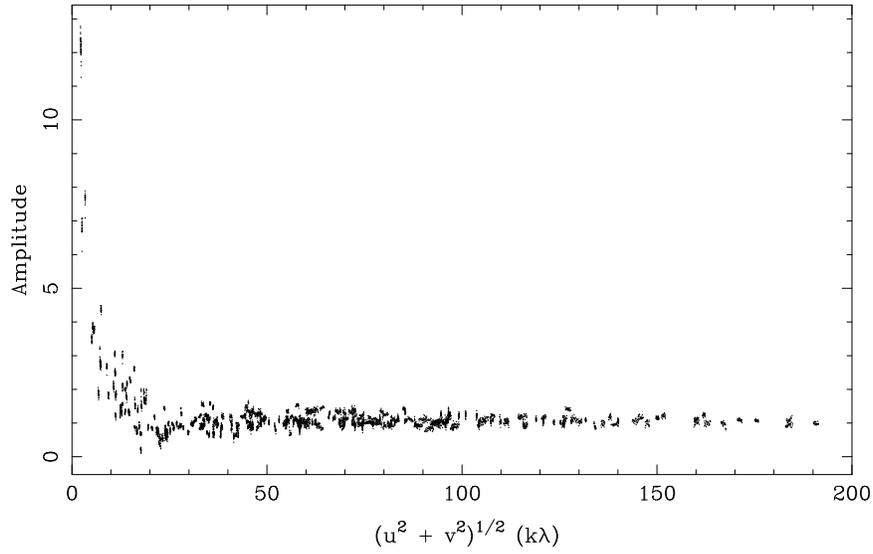}
\caption{Plot of amplitude (in Jy) versus $u,v$ distance for Sgr~A*.
This data was taken at a wavelength of 1.3~cm on 19 December 2002 in the
C Array of the VLA.  Contributions from Sgr A West are negligible on
baselines longer than $\sim40$~k$\lambda$.  \label{uv}}
\end{figure}

The flux density of Sgr~A* was measured in the {\it u,v} domain.  In
determining the flux density of Sgr~A*, it is important to avoid
contributions from the extended source, Sgr A West, that surrounds the
supermassive black hole.  In Figure \ref{uv}, a plot of amplitude
versus {\it u,v} distance for Sgr~A* shows the contribution from Sgr A
West on baselines $\simlt40$~k$\lambda$ at 1.3~cm (angular scales
$\simgt5''$).  For observations made in the largest configurations of
the VLA, only baselines longer than 100~k$\lambda$ (corresponding to a
$2''$ resolution) were used in the flux density calculation.  For more
compact arrays, it was necessary to decrease the minimum allowed {\it
u,v} distance for the longer wavelength data.  At 2.0~cm in C Array
and 1.3~cm in D Array, the minimum {\it u,v} distance was
60~k$\lambda$.  In D Array, baselines longer than 40~k$\lambda$ were
used for 2.0~cm data.  Finally, an initial estimate of the uncertainty
in the flux density of Sgr~A* was calculated assuming that Sgr~A* has
the same fractional error in flux density ($\frac{\Delta S}{S}$) as
the phase calibrators.

\subsection{Correcting for Systematic Scaling Errors in the Flux Density Calibration\label{sdo.sec}}

\begin{figure}
%\epsscale{0.8}
%\plotone{Herrnstein.fig2.eps}
\centering
\includegraphics[width=4.5in, angle=90]{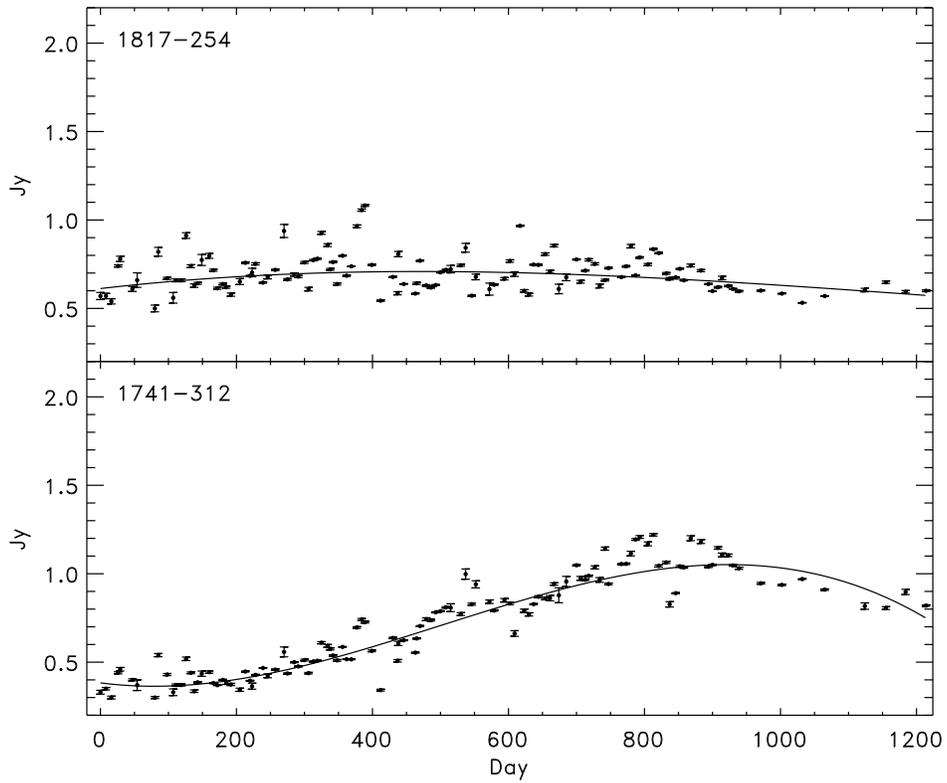}
\caption{Flux density of 1817--254 {\it (top)} and 1741--312 {\it
(bottom)} at 1.3~cm after calibration in {\it AIPS}.  The best cubic fit to
the data is overlaid. The vertical axis is chosen
to match Figure \ref{all}.\label{calfit}}
\end{figure}

We present 1.3~cm flux densities for 1817--254 and 1741--312 after
initial calibration in {\it AIPS} in Figure \ref{calfit}.  Variability
in flux density of the two calibrators is due to a long-term drift in
flux density, short-term variability intrinsic to the source, and
slight differences in the absolute flux density scale between epochs.
In the following paragraphs, we present the method used to calculate
and correct for systematic changes in the flux density scale between
epochs.  Because the intrinsic short-term variability of the
calibrators is not known, we assume that calibrators have only
long-term drifts in flux density.  Short-term variability, which is
highly correlated between the two calibrators, is assumed to result
from changes in the flux density scale between epochs.  These scale
factors are quantified and removed from the data.  If any intrinsic,
short-term variability of the calibrators exists, it is accounted for
in the calculation of the final uncertainties in these scale factors
presented at the end of this section.

\subsubsection{Calculation of Gain Adjustment Factors \label{gaf.sec}}

To estimate the long-term drift in flux density, a cubic fit has been
made to the light curve of each calibrator (see Figure \ref{calfit}).
The parameters of the best fit model for 1817--254 and 1741--312 at
each wavelength are given in Table \ref{cubtable}.  At all three
wavelengths, 1741--312 is consistent with a cubic profile.  Calibrator
1817--254, however, shows little long-term change in flux over our
entire monitoring campaign.  This is reflected by the relatively small
values of the linear, quadratic, and cubic terms for this calibrator
(see Table \ref{cubtable}).

\begin{figure}
%\plotone{Herrnstein.fig3.eps}
%\epsscale{0.8}
\centering
\includegraphics[width=4.2in, angle=90]{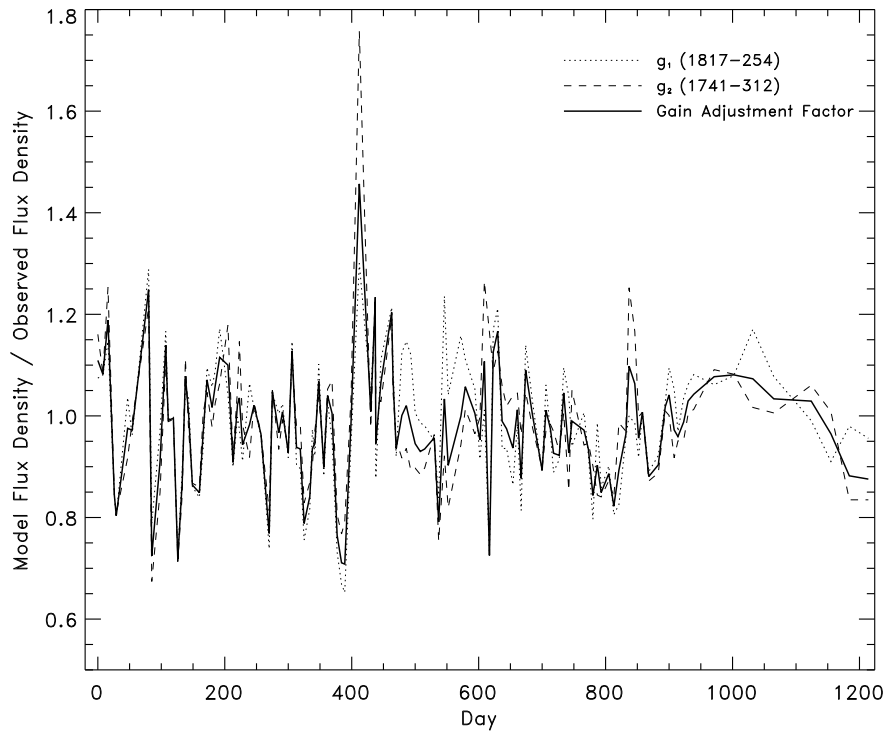}
\caption{Ratio of model flux density derived from the cubic fit to the
observed flux density at 1.3~cm for 1817--254 ($g_1$) and 1741--312
($g_2$).  The gain adjustment factor (GAF), calculated as the weighted
mean of $g_1$ and $g_2$ at each epoch, is overlaid as a solid line
(see \S \ref{gaf.sec}). \label{sdo}}
\end{figure}

Both phase calibrators show additional short-term variability that is
not accounted for by the cubic model.  For each epoch, the ratio of
the model flux density to the observed flux density is calculated for
each calibrator.  If we define calibrator 1 as 1817--254 and
calibrator 2 as 1741--312, then these ratios are given by \bq
g_1=S_1/S'_1~~{\rm and}~~g_2=S_2/S'_2~, \eq
\noindent where $S_i$ is the model flux density from the cubic fit and
$S'_i$ is the observed flux density.  The uncertainty in $g_i$ is
calculated as the quadrature sum of the uncertainty in the observed
flux density and the uncertainty in the model flux density from the
cubic fit.  Figure \ref{sdo} shows a plot of $g_1$ and $g_2$ for every
epoch during the monitoring campaign.  Intrinsic short-term
variability of the two calibrators should result in uncorrelated
values of $g_1$ and $g_2$.  Systematic errors in the absolute flux
density scale, however, will appear as a correlation between the two
ratios in Figure \ref{sdo}. The values of $g_1$ and $g_2$ are
well-correlated, indicating that systematic errors in the absolute
flux density calibration are a significant component of the calibrator
variability.

We calculate a gain adjustment factor (GAF) as the weighted average of
$g_1$ and $g_2$:
\bq
g=\frac{g_1/\sigma_{g_1}^2 + g_2/\sigma_{g_2}^2}{1/\sigma_{g_1}^2 + 1/\sigma_{g_2}^2}~. 
\eq
\noindent The gain adjustment factors track the systematic change in
the flux density scale with time.  The GAF ``light curve'' at 1.3~cm
is shown as the solid line in Figure \ref{sdo}.  At all three
wavelengths, the average value of the GAF is close to 1, indicating
that there is no systematic underestimate or overestimate of the flux
densities.  However, the typical size of the fractional correction
increases towards shorter wavelengths.  The standard deviation of $g$
is 0.07, 0.11, and 0.17 at 2.0, 1.3, and 0.7~cm, respectively.

\subsubsection{Estimation of Uncertainties \label{unc.sec}}

Typical errors in flux density reported by {\it AIPS} range from
1--3\%.  The fact that the values of $g_1$ and $g_2$ are not identical
to within these errors (see e.g. day 486--515 in Figure \ref{sdo})
indicates that {\it AIPS} has underestimated the true uncertainties in
the flux density measurements and/or there is significant intrinsic
variability of one or both of the phase calibrators.  The errors
reported by {\it AIPS} are statistical and do not include
contributions from systematic errors.  Due to these systematic errors
(e.g. pointing errors) the flux density calibration at any given epoch
is expected to be accurate to only $5-10$\%.  The intrinsic
variability of the two phase calibrators is not known.  Therefore,
while both an underestimation of the errors by {\it AIPS} and
intrinsic source variability probably contribute to the difference
between $g_1$ and $g_2$, we take the conservative approach and assume
that all of the residual variability observed is due to systematic
errors in the calibration.

\begin{figure}
%\plotone{Herrnstein.fig4.eps}
%\epsscale{0.8}
\centering
\includegraphics[width=4.2in, angle=90]{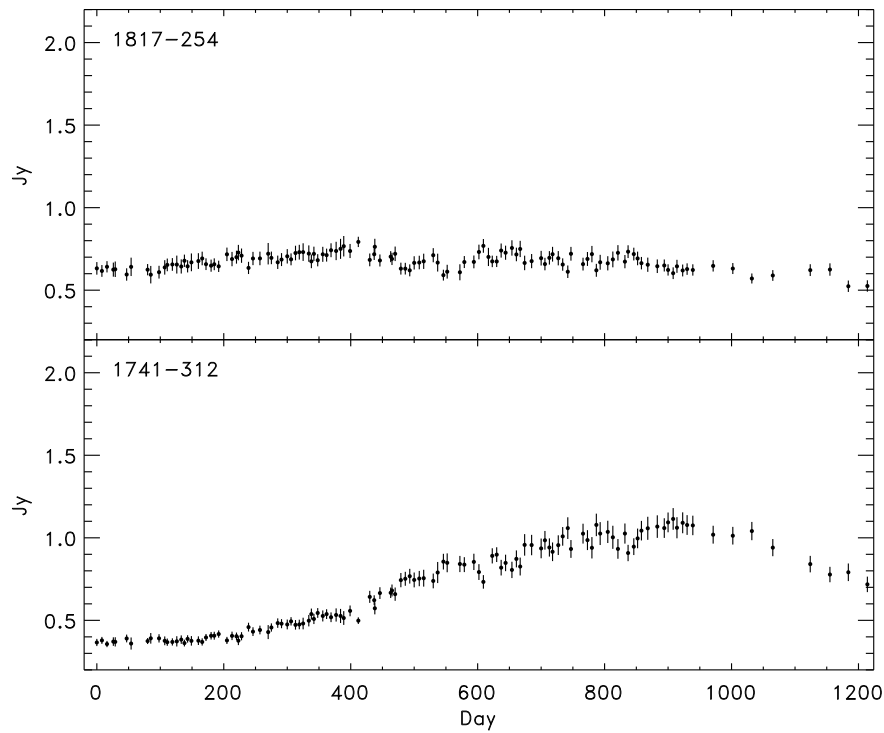}
\caption{Flux density of 1817--254 {\it (top)} and 1741--312 {\it
(bottom)} at 1.3~cm after application of the gain adjustment factors.
Error bars include the original error and the additional fractional
error discussed in Section \ref{unc.sec}. The vertical axis is chosen
to match Figure \ref{all}.  \label{calvar}}
\end{figure}

Assuming that the phase calibrators have only slow drifts in flux
density and no short-term variability to within the noise, the
difference between $g_1$ and $g_2$ in Figure \ref{sdo} can be used to
estimate the true rms noise, $\sigma'$, in the flux density
measurements.  This assumption is reasonable for both of our
calibrators (see Figure \ref{calvar}).  An additional error,
calculated as some fraction, $f$, of the source flux density, is added
in quadrature to the initial error reported by {\it AIPS}.  For epoch
$i$, \bq \sigma'(i)=\sqrt{(fS_\nu(i))^2 + \sigma_{AIPS}(i)^2}~. \eq
\noindent The value of $f$ is chosen such that the fit of the GAFs to
the residual variability of the two calibrators has a reduced
chi-squared of one.  The values of $f$ for our data are 0.059, 0.057,
and 0.078 for 2.0, 1.3, and 0.7~cm, respectively.  Final errors in the
flux density measurements for Sgr~A* are typically 6.1\%, 6.2\%, and
9.2\% for 2.0, 1.3, and 0.7~cm, respectively.

A gain adjustment factor could not be determined for the observation
on 16 June 2003.  In this case, we calculate the uncertainty in the
flux density measurement as the quadrature sum of the mean uncertainty
in the fully-calibrated flux densities of Sgr~A* and the mean size of
the gain adjustment factors.

\subsubsection{Final Light Curves}

Fully-calibrated flux densities for 1817--254 and 1741--312 are
calculated by multiplying the observed flux density by the gain
adjustment factor.  The resulting light curves at 1.3~cm are shown in
Figure \ref{calvar}.  After application of the GAF, the flux densities
of both phase calibrators show only long-term drifts to within the
uncertainties.  For each epoch, the final flux density and associated
uncertainty for Sgr A* is calculated in the same way as for the phase
calibrators and is equal to $gS'_*$.

\section{Results \label{res}}

\begin{figure}
%\plotone{Herrnstein.fig5.eps}
%\epsscale{0.8}
\centering
\includegraphics[width=4.2in, angle=90]{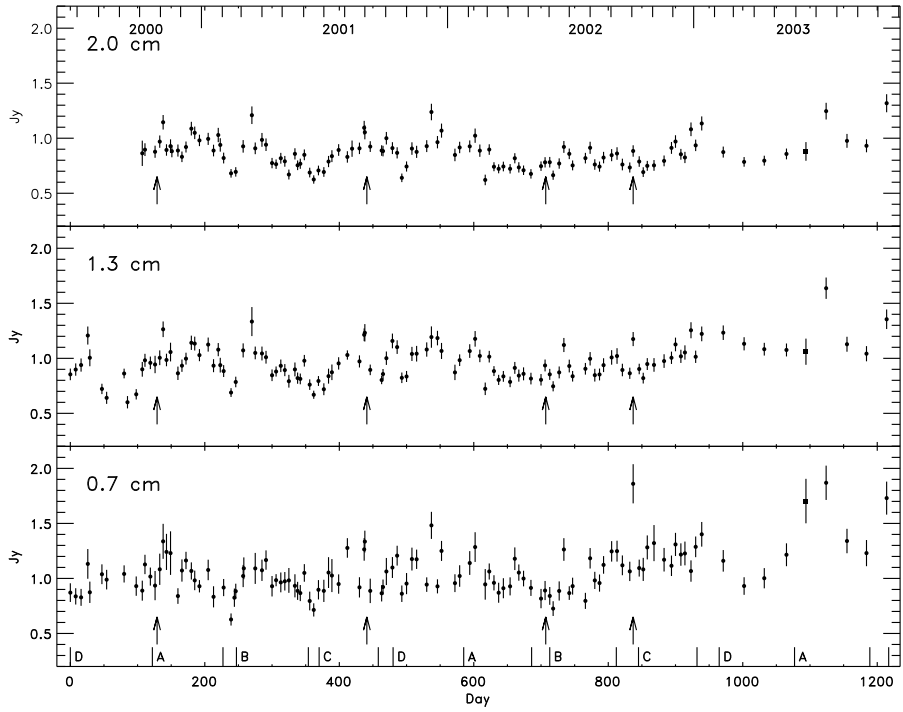}
\caption{Intrinsic variability of Sgr~A* at 2.0, 1.3, and 0.7~cm.  The
corresponding date is labeled in the top panel and the VLA Array is
labeled in the bottom panel.  Hybrid configurations are not
labeled. Arrows mark four large X-ray flares (see
\S\ref{intro}). \label{all}}
\end{figure}

Figure \ref{all} shows the radio light curves of Sgr~A* at 2.0, 1.3,
and 0.7~cm.  The fully calibrated flux densities and associated errors
are listed in Table \ref{table} and in the online version of this
article.  Table \ref{sumtable} summarizes the characteristics of the
variability of Sgr~A* at each wavelength.

Table \ref{array_tab} shows the mean and standard deviation of the
flux density of Sgr~A* as a function of the VLA array in which the
observations were made.  VLA configurations are changed roughly every
four months, with slight variations due to subscription and
scheduling.  The 16-month cycle begins in the largest configuration
(A) and moves to progressively smaller configurations, ending in the D
configuration.  As the antennas are moved to a new configuration,
roughly three weeks are spent in a hybrid array, during which the
northern arm of the array remains in the previous (more extended)
configuration and the eastern and western arms are in the new (more
compact) configuration.  These hybrid arrays are used primarily for
sources at low declinations.  At the end of the cycle, roughly three
weeks are spent moving the antennas from the D configuration back to
the A configuration.  Because the scheduling of configuration changes
and the time spent in each configuration can vary by as much as 2--3
weeks, we do not expect the changes in configuration to produce
periodic signatures in our data.  This point will be discussed in more
detail when the data are analyzed for periodicities, but we can
conclude from Table \ref{array_tab} that no trends in flux density or
variability as a function of observing array are apparent.

\begin{figure}
%\plotone{Herrnstein.fig6.eps}
%\epsscale{0.8}
\centering
\includegraphics[width=4.2in, angle=90]{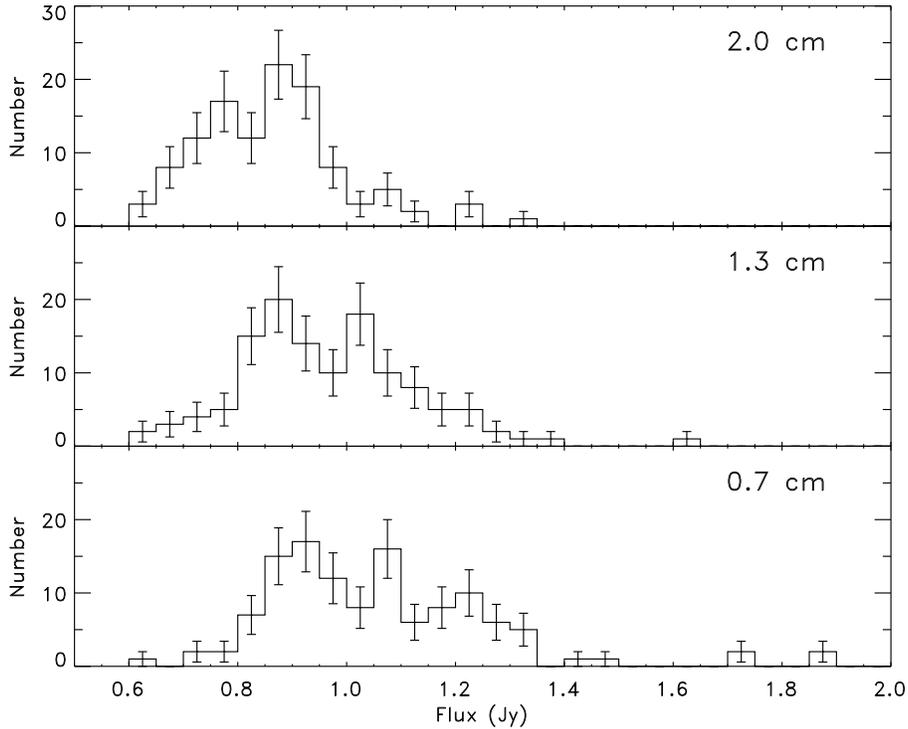}
\caption{Histogram of flux densities at each wavelength.  The bin size is 50~mJy.  At all three wavelengths, the distribution of flux densities shows evidence for two distinct peaks.  A high flux density tail is apparent at 1.3 and 0.7~cm, reflecting the increased variability towards shorter wavelengths.  \label{hist}}
\end{figure}

Figure \ref{hist} plots a histogram of the flux densities at each
wavelength using a bin size of 0.05~Jy.  This bin size is roughly
equal to the mean uncertainty in the flux density of Sgr~A* at 2.0 and
1.3~cm.  Uncertainties in the distribution are calculated as the
square root of the number of points in each bin.  At all three
wavelengths, there appears to be a tendency towards a bimodal
distribution of flux densities.  If the bimodal distribution is real,
it may reflect two different states of accretion onto the supermassive
black hole.

Significant variability in the flux density is observed at all three
wavelengths (see Figure \ref{all}).  The reduced chi-squared of a fit
of a constant flux density to the observed light curve was calculated
at each wavelength.  The reduced chi-squared is equal to 5.7 at
2.0~cm, 6.7 at 1.3~cm, and 3.9 at 0.7~cm.  The similar shape of the
three light curves indicates that the flux densities at centimeter
wavelengths are also highly correlated.  Figure \ref{hist} shows high
flux density tails in the distributions at 1.3 and 0.7~cm.  These
tails reflect the increased variability towards shorter wavelengths,
consistent with the results of \citet{zha92}.  The largest variability
in flux densities was observed at 0.7~cm, where the flux density of
Sgr~A* varied from a minimum value of 0.63\p0.06~Jy to a maximum value
of 1.86\p0.16~Jy.  The flux density of Sgr~A* has a standard deviation
during our monitoring campaign of 0.13, 0.16, and 0.21~Jy at 2.0, 1.3,
and 0.7~cm, respectively.  

A comparison of our light curves to the regularly sampled data from
ZBG01 indicates that Sgr~A* may have been in a relatively quiescent
state from 2000.5 to 2003.8. Between 1990.1 and 1991.5, ZBG01
monitored the flux density of Sgr~A* with sampling intervals ranging
from 1 to 28 days.  These finely sampled data form a small subset of
the entire 20-year ZBG01 dataset.  During the 1.4 years of regular
observations, two outbursts in which the 1.3~cm flux density of Sgr~A*
exceeded twice the mean value were observed.  At least three
additional outbursts of smaller amplitude were also observed.  

Strong outbursts appear to be less frequent in the recent monitoring
data.  We observed no outbursts in which the flux density of Sgr A*
doubled.  Only one outburst at 2.0~cm had an amplitude $>50$\% higher
than the mean flux density (0.834\p0.005~Jy).  This outburst occurred
on 14 October 2003 and had a flux density of 1.32\p0.08~Jy.  At
1.3~cm, the largest outburst was a $4.5\sigma$ event on 16 July 2003
(day 1124).  On this date, the flux density of Sgr~A* was
1.64\p0.10~Jy, 77\% higher than the mean value during our monitoring
campaign (0.926\p0.005~Jy).  At 0.7~cm, we detected two 4.1$\sigma$
outbursts in which the flux density of Sgr~A* was $\sim86$\% higher
than the mean value (1.001\p0.008~Jy).  An outburst with a flux
density of 1.86\p0.18~Jy occurred on 3 October 2002 (day 837),
followed by a second outburst on 16 July 2003 (day 1124) with a flux
density of 1.87\p0.16~Jy.  Although there are relatively few strong
outbursts overall, many of the largest outbursts at all three
wavelengths occurred within the past year.  This result suggests that
Sgr~A* may have become more active beginning in mid 2003.

Initial comparison of the radio light curves with X-ray observations
of Sgr~A* indicates that there may be a correlation between strong
X-ray flares and increases in the flux density at centimeter
wavelengths.  The dates of the four strong X-ray flares discussed in
\S\ref{intro} are marked by arrows in Figure \ref{all}.  In
particular, VLA observations made on 03 October 2003, just 13.5 hours
after the onset of the factor of 160 flare, show highly elevated flux
densities at all three wavelengths.  The 0.7~cm flux density for this
date ($1.86\pm0.18$~Jy) was one of the two largest flux densities
measured during our entire campaign.  This date also marked the
steepest measured spectral index (0.70\p0.10, see \S \ref{alpha}).
Correlated X-ray and radio variability may provide clues to the
underlying physical processes near the supermassive black hole.
Comparison of the radio and X-ray light curves, including a discussion
of the relationship between the radio and X-ray events on 03 October
2003, is presented in detail in \citet{zha04}.

Recent observations of Sgr~A* with the Sub-Millimeter Array (SMA)
indicate that sub-millimeter and centimeter flux densities are also
correlated.  The SMA (operated jointly by the Smithsonian
Astrophysical Observatory and the Academica Sinica Institute for
Astronomy and Astrophysics) is the first interferometric array to
work full-time at wavelengths $\simlt 1$~mm.  The high resolution of
this interferometer will allow Sgr~A* to be separated from the
surrounding thermal dust emission from Sgr A West.  Although still
under construction at the time, monitoring of the flux density of
Sgr~A* at 1.3~mm began in 2001 \citep{zha03}.  Between March 2001 and
July 2002, the flux density of Sgr~A* was measured at 24 epochs with a
resolution of $2-10''$.  The sampling of this dataset is too sparse to
search for periodicities or determine the overall characteristics of
the light curve, but a comparison with the VLA monitoring data
indicates that the brightest flux densities at 1.3~mm occurred at
times when the flux densities at centimeter wavelengths were also high
\citep[see Figure 2]{zha03}.  In the future, more frequent SMA
observations of Sgr~A* at 1.3 and 0.87~mm will enable a comparison of
the light curves at millimeter and centimeter wavelengths in more
detail.

Finally, it has also been suggested that there may be a correlation
between the radio flux density of Sgr A* and the closest approaches of
orbiting stars.  \citet{loeb04} has predicted that the flux density of
Sgr A* will vary on time scales $>1$~month due to fluctuations in the
mass accretion rate as stars approach and recede from Sgr A*.  The
predicted magnitude of an outburst as well as the time delay between
closest approach and the outburst are highly uncertain. \citet{ghe03}
have identified three periapse dates for massive stars in the Galactic
Center: SO--19 (1995.639), SO--16 (2000.243) and SO--2 (2002.335).
Only the periapse for SO--2 occurred during our monitoring campaign
(near day 685).  We see no evidence in the light curves for an
outburst occurring within 100 days of the SO--2 periapse.  There is
also no evidence for an increase in the mean flux density or a change
in the spectral index on this time scale.  Therefore, we conclude that
the emission from Sgr A* at centimeter wavelengths is not
significantly affected on time-scales $\simgt1$~week by the close
approach of massive stars.

\subsection{Spectral Index - Flux Density Correlation\label{alpha}}
\begin{figure}
%\plotone{Herrnstein.fig7.eps}
%\epsscale{0.8}
\centering
\includegraphics[width=4.2in, angle=90]{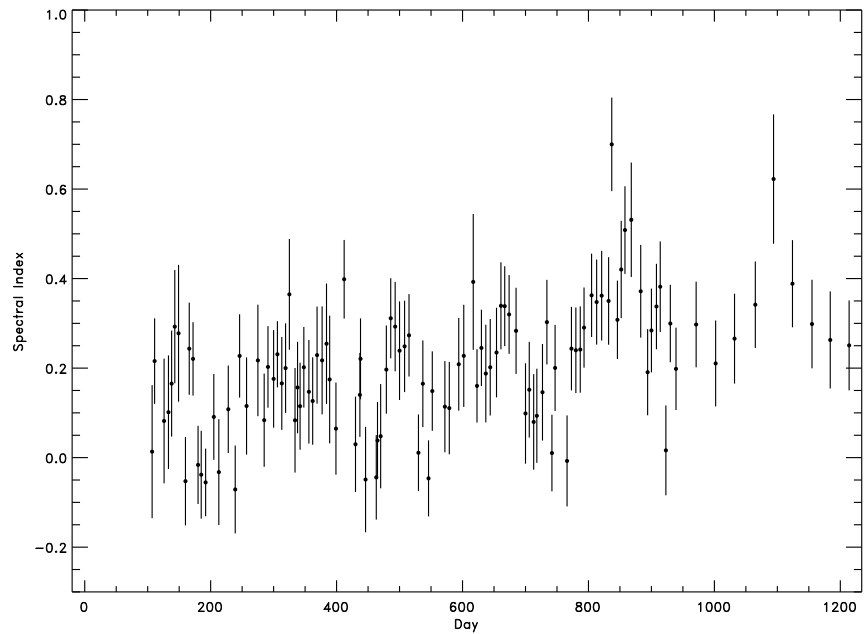}
\caption{Spectral index $\alpha$ ($S_\nu\propto\nu^\alpha$) as a function of day for every epoch in which flux density measurements were obtained at all three wavelengths.   \label{index}}
\end{figure}

\begin{figure}
%\plotone{Herrnstein.fig8.eps}
%\epsscale{0.8}
\centering
\includegraphics[width=4.2in, angle=90]{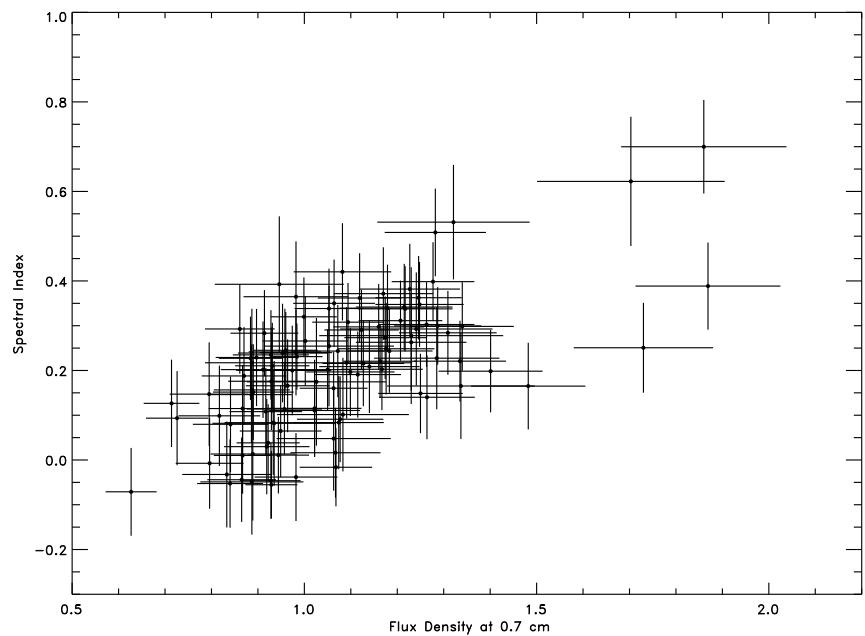}
\caption{Spectral index $\alpha$ ($S_\nu\propto\nu^\alpha$) as a function of
flux density at 0.7~cm.  The spectral index becomes steeper during
outbursts of Sgr~A*. \label{corr}}
\end{figure}

The spectral index $\alpha$ (defined as $S_\nu\propto\nu^\alpha$) can
be calculated from the measured flux densities at 2.0, 1.3, and
0.7~cm.  In Figure \ref{index}, we plot the calculated spectral index
at every epoch in which the flux density was determined at all three
wavelengths.  The calculated values for $\alpha$ are also listed in
Table \ref{table}.  During the monitoring campaign, the spectral index
was observed to vary between $-0.07\pm0.12$ and $0.70\pm0.10$.  The
standard deviation of the calculated spectral indices is 0.14.  In
Figure \ref{corr}, the spectral index is plotted as a function of the
observed flux density at 0.7~cm ($S_{0.7}$).  The spectral index
depends linearly on $S_{0.7}$, with a best fit of
$\alpha=-0.41(\pm0.04)+0.57(\pm0.04)S_{0.7}$.  The tendency towards
steeper spectral index during outburst states is consistent with recent
observations in the sub-millimeter \citep{zha03}.

Increased fractional variability towards shorter centimeter
wavelengths has been used to suggest that the observed variability of
Sgr~A* is intrinsic to the source \citep{zha92}.  The linear
dependence of spectral index on 0.7~cm flux density strongly favors a
model in which the observed variability is intrinsic to Sgr~A* and not
the result of interstellar scattering.  If the flux density
variability is caused by an outburst with a self-absorbed synchrotron
spectrum (with $\alpha_{{\rm sync}} \approx 2$), then the flux density
and the spectral index will rise together (e.g., \citet{marscher85}).
For our data, the peak spectral index will occur when the
self-absorption frequency is equal to 43~GHz (0.7~cm).  At this point,
the flux density at this frequency will also be at a maximum.  As the
self-absorption frequency moves to lower frequencies, both the
spectral index and the 0.7~cm flux density will decrease, producing a
correlation between spectral index and flux density.  If the optically
thin spectral index is $\ll0$, then the spectral index will quickly
become negative when the self-absorption frequency moves to lower
frequencies.  Because the measured spectral indices for Sgr~A* are
almost all greater than zero, the optically thin spectral index must
be reasonably flat ($\alpha_{{\rm thin}}\simgt-0.5$).

Although we believe the above interpretation represents the most
likely scenario, it is not unique.  Due to the complexity and
peculiarity of the Galactic Center scattering screen, interstellar
scattering cannot be ruled out as the source of the observed
variability.  However, a positive correlation between flux density and
spectral index is not generally expected for interstellar
scintillation.  In simple models of interstellar scattering, the
modulation index of the flux density decreases with frequency in the
strong scattering regime \citep{rickett90}.  This would lead to an
anti-correlation between spectral index and flux density. Furthermore,
the time scale for diffractive scintillation for Sgr~A* is $<100$~s
while the time scale for refractive scintillation is $3\times10^6$~s,
assuming relative velocities for the Earth and the scattering medium
of $\sim100$~km~s$^{-1}$ \citep{rickett90}.  We detect strong
variability on a time scale that falls well in between these
characteristic times.  Refractive scintillation appears to be only
relevant for velocities $>1000$~km~s$^{-1}$, which would likely be
associated with stellar winds in the central parsec or the accretion
flow itself.  To match the correlation (and the increased variability
at millimeter and sub-millimeter wavelengths), the turbulence must
increase substantially with decreasing scale size in the scattering
medium.

The mean spectral index of Sgr~A* can be calculated using the average
flux densities at all three wavelengths (see Table \ref{sumtable}).
The best fit to the mean spectral index is $\alpha=0.20\pm0.01$.
There is strong evidence for a break in the spectrum of Sgr~A*
resulting in an excess in the flux density observed at 1~mm
\citep{zyl92,ser97,fal98,zha03}.  \citet{fal98} calculate a spectral
index of $\alpha=0.17$ at wavelengths longer than 2~cm, but find a
spectral index of $\alpha=0.3$ at wavelengths of 2~cm and shorter.
This measurement is based on a single epoch of data, and it is not
unexpected given the observed standard deviation of the spectral index
of 0.14 during our monitoring campaign.  Large changes in the spectral
index of Sgr~A* can be seen in our data, and many epochs have spectral
indices $\ge0.3$ (see Table \ref{table}).  Our data, however, suggest
that, on average, the break occurs at wavelengths shorter than 7~mm.
\citet{zha03} also find a spectral index of $0.1\pm0.1$ between 2.0~cm
and 3~mm using one epoch of VLA data at 2.0, 1.3, and 0.7~cm and a
measurement of the flux density at 3~mm made by \citet{tsu02}.
Between 3~mm and 0.87~mm, the spectral index rises to 0.25, indicating
that the break in the spectrum occurs near 3~mm.  It appears that,
like the spectral index, the wavelength of the break in the spectrum
is likely variable.  Coordinated observations from centimeter to
sub-millimeter wavelengths will be necessary to determine the precise
characteristics of the break in the spectrum of Sgr~A*.

A prolonged period of increased spectral index and 0.7~cm flux density
appeared to begin around 1 September 2002 and has persisted throughout
the remainder of the monitoring campaign.  This period also
corresponds to the time of an observed increase in variability of
Sgr~A* (see \S \ref{res}).  Before 1 September 2002 (day 805), the
weighted mean spectral index was 0.16\p0.01 with a standard deviation
of 0.11.  The mean flux density at 0.7~cm was 0.94\p0.01~Jy.  After 1
September, the weighted mean of the spectral index and flux density at
0.7~cm increased to 0.33\p0.02 and 1.20\p0.02~Jy, respectively.  The
spectral index is also more variable during this time and has a
standard deviation of 0.14.

\subsection{Time Delay}
\begin{figure}
%\plotone{Herrnstein.fig9.eps}
%\epsscale{0.8}
\centering
\includegraphics[width=4.2in]{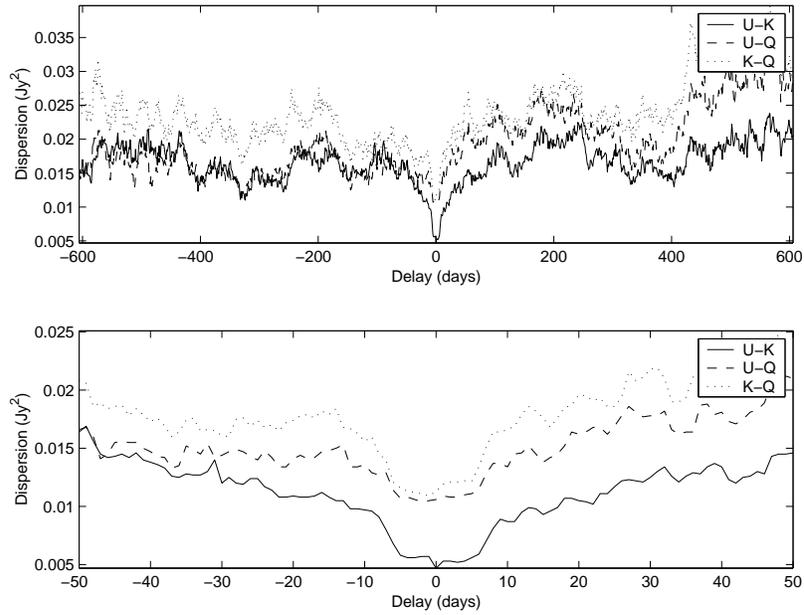}
\caption{The dispersion as a function of time delay between pairs of
light curves at the three different frequencies (U=2.0~cm, K=1.3~cm,
and Q=0.7~cm).  There is a clear minimum in the dispersion at zero
delay for each pair.  \label{disp}}
\end{figure}

The time sampling of our monitoring campaign makes our data sensitive
to delays between wavelengths roughly greater than one week.  The time
delay between the flux density at the three different frequencies is
estimated using the non-parametric method of Pelt et al. (1994).  This
method is used to calculate the dispersion between two irregularly
sampled light curves by searching for a magnification and a shift in
time of the two curves.  The dispersion is essentially the mean square
difference between the flux density at two frequencies on a given time
scale.  Figure \ref{disp} shows the dispersion between the light
curves for each possible pair of wavelengths.  The results strongly
favor no delay between any pair of the three frequencies.  A minimum
is found in the dispersion for delays less than $\sim 5$ days, which
corresponds to the minimum sampling interval in our data.  A
monitoring campaign with sampling intervals on daily to hourly time
scales will be necessary to determine time delays between wavelengths.

\begin{figure}
%\plotone{Herrnstein.fig10.eps}
%\epsscale{0.8}
\centering
\includegraphics[width=4.2in]{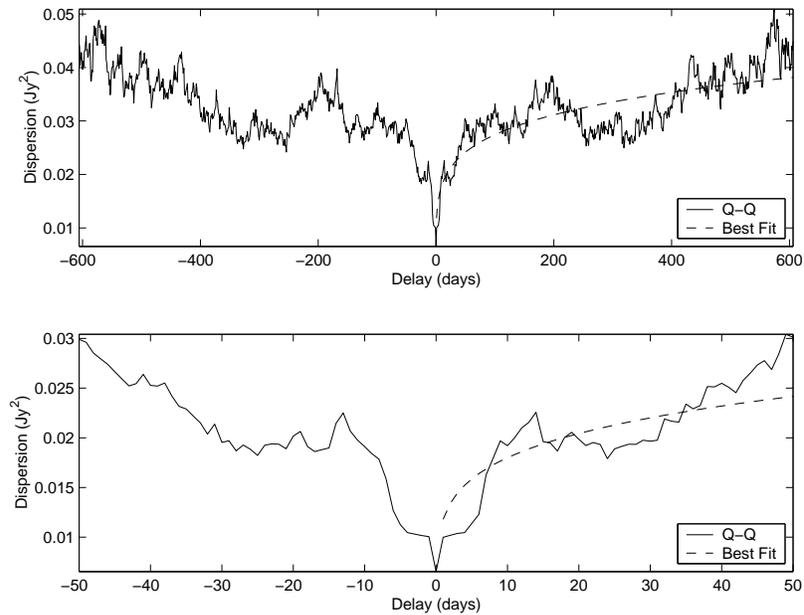}
\caption{The auto-dispersion as a function of time delay
for the 0.7~cm light curve.  A power-law fit of $D^2 \propto t^{0.2}$
is also shown. \label{dispq}}
\end{figure}

We also calculate the auto-dispersion of the data by comparing each
light curve with itself.  This calculation enables us to estimate the
characteristic change in flux density with time.  Results for all
three wavelengths are similar, and the auto-dispersion for the 0.7~cm
light curve is shown in Figure \ref{dispq}.  The auto-dispersion grows
slowly with time ($D^2 \propto t^{0.2}$).  The weak dependence on time
is consistent with the fairly static mean flux density that has been
observed since the discovery of Sgr~A* (e.g., ZBG01; \citet{bower02}).

\section{Conclusions}

In this paper, we have presented results from a 3.3-year campaign to
monitor the flux density of Sgr~A* at centimeter wavelengths using the
VLA.  The largest amplitude variations are observed at 0.7~cm,
consistent with variability increasing towards shorter wavelengths
(ZBG01).  Overall, however, Sgr~A* appears to be more quiescent than
during previous monitoring campaigns.  The spectral index of Sgr~A*
appears to be strongly correlated with the 0.7~cm flux density.  This
result strongly favors an emission mechanism in which outbursts are
intrinsic to Sgr~A* and are not the result of interstellar
scintillation.  Regular monitoring of Sgr~A* at the VLA will continue
through at least June 2004 with monthly observations at 2.0, 1.3, and
0.7~cm.  These additional data will be useful in detecting periods
longer than 100 days.

Much of the activity of Sgr~A* appears to take place on time scales
$\simgt1$~hr and less than the time resolution of our data (8 days).
However, the monitoring campaigns in the radio and sub-millimeter have
only minimally probed these time scales.  To date, hourly time scales
have only been systematically probed in X-rays and the infrared.  In
both cases, there is significant variability on time scales of hours.
Monitoring of the centimeter flux density on these short time scales
will be necessary to determine the duration and shape of outbursts as
well as detect any time lag between wavelengths.  Additional
coordinated multi-wavelength campaigns specifically aimed at probing
these time scales should also be implemented.  Only with simultaneous
coverage and fine sampling will we be able to ascertain the
relationship between radio outbursts and X-ray flares and constrain
the emission mechanism for Sgr~A*.

\acknowledgements{The authors would like to thank C. Fassnacht for
help with the initial data calibration and sharing his code to
calculate the dispersion between light curves.  We also thank
C. Chandler for the model image of 3C286 used in the data calibration
and J. Herrnstein for useful discussions.}
%The National Radio Astronomy Observatory is a
%facility of the National Science Foundation operated under cooperative
%agreement by Associated Universities, Inc.}

\bibliographystyle{apj} 
\bibliography{Herrnstein}

\newpage

\newpage
%%%%%%%%%%%%%%%%%%%%% BEGIN TABLE 1 %%%%%%%%%%%%%%%%%%%%%%%
\begin{deluxetable}{l l r r r r } 
\tablewidth{0pt}
\tablecaption{Cubic Fit Parameters of Phase Calibrator \label{cubtable}}
\tablehead{ 
\colhead{Phase} & \colhead{Observing} & \colhead{a$_0$} & \colhead{a$_1$} & \colhead{a$_2$} & \colhead{a$_3$} \\
\colhead{Calibrator} & \colhead{Band} & \colhead{(mJy)} & \colhead{(mJy yr$^{-1}$)} & \colhead{(mJy yr$^{-2}$)} & \colhead{(mJy yr$^{-3}$)}}
\startdata
1817--254\\
 & 2.0~cm & 794\p~6 &   47\p10 & --80\p10 & 17\p~8 \\
 & 1.3~cm & 612\p~5 & 162\p~8 & --78\p~6 & 8\p~5 \\
 & 0.7~cm & 388\p~5 &  5\p~8 & --82\p~6 & --26\p~5 \\
1741--312\\
 & 2.0~cm & 540\p~5 & --444\p10 & 542\p~7 & --117\p~7 \\
 & 1.3~cm & 383\p~5 & --182\p~9 & 459\p~5 & --112\p~5  \\
 & 0.7~cm & 287\p~6 & --163\p10 & 419\p~5 & --105\p~5 \\
\enddata
\tablecomments{The cubic fit is of the form $S_\nu=a_0+a_1x+a_2x^2+a_3x^3$, where $x$ is measured in years and $x=0$ corresponds to day 0 in Table \ref{table}.}
\end{deluxetable}

%%%%%%%%%%%%%%%%%%%%% BEGIN TABLE 2 %%%%%%%%%%%%%%%%%%%%%%%

\newpage
\begin{deluxetable}{r r c c c r} 
\tabletypesize{\small} 
\tablewidth{0pt} 
\tablecaption{Calibrated Flux Density of Sgr~A*\label{table}} 
\tablehead{ \colhead{Date} &
\colhead{Project} & \colhead{2.0~cm} & \colhead{1.3~cm} &
\colhead{0.7~cm} & \colhead{Spectral} \\ & \colhead{day\tablenotemark{\dagger}} &
\colhead{(Jy)} & \colhead{(Jy)} & \colhead{(Jy)} & \colhead{Index} }
\startdata
\input{Herrnstein.tab2.dat}
\tablenotetext{\dagger}{Project day 0 is 21 June 2000.  Days refer to
LST days such that observations occurred at roughly the same time each
day.}  \enddata
\end{deluxetable}

%%%%%%%%%%%%%%%%%%%%% END TABLE 2 %%%%%%%%%%%%%%%%%%%%%%%

%%%%%%%%%%%%%%%%%%%%% BEGIN TABLE 3 %%%%%%%%%%%%%%%%%%%%%%%
\begin{deluxetable}{ccccccc}
\tabletypesize{\small}
\tablewidth{0pt}
\tablecaption{Parameters for the Flux Density Variability of Sgr~A* \label{sumtable}}
\tablehead{ 
\colhead{Wavelength} & \colhead{Number} & \colhead{$\langle S_\nu\rangle$} & \colhead{$S_{min}$} & \colhead{$S_{max}$} & \colhead{$\sigma$} & \colhead{$\chi_r^2$ of Fit to} \\
\colhead{(cm)} & \colhead{of Epochs} & \colhead{(Jy)} & \colhead{(Jy)} & \colhead{(Jy)} & \colhead{(Jy)} &\colhead{Constant $S_\nu$}}
\startdata
2.0 & 115 & 0.834\p0.005 & 0.62\p0.05 & 1.32\p0.08 & 0.13 & 5.7 \\
1.3 & 124 & 0.926\p0.005 & 0.60\p0.06 & 1.64\p0.10 & 0.16 & 6.7 \\
0.7 & 121 & 1.001\p0.008 & 0.63\p0.06 & 1.87\p0.16 & 0.21 & 3.9 \\
\enddata
\tablecomments{The error in the mean flux density is calculated as the weighted error from all measurements.  The standard deviation of the measured flux densities at each wavelength is given as $\sigma$.}
%\tablenotetext{\dagger{The variable $f$ defines the additional fractional error in the flux density, $\sigma_f=f~S_\nu(i)$, that must be added to the initial error reported by AIPS to ensure that the fit of the GAFs to the residual variability of the two phase calibrators has a reduced chi-squared value of one (see \S \ref{unc.sec}).}
\end{deluxetable}

%%%%%%%%%%%%%%%%%%%%% END TABLE 3 %%%%%%%%%%%%%%%%%%%%%%%

%%%%%%%%%%%%%%%%%%%%% BEGIN TABLE 4 %%%%%%%%%%%%%%%%%%%%%%%
\newpage
\begin{deluxetable}{cccccccccc}
\tabletypesize{\small}
\tablewidth{0pt}
\tablecaption{Flux Density as a Function of Array \label{array_tab}}
\tablehead{ 
\colhead{} & \multicolumn{3}{c}{\underline{\phantom{\hspace{2em}}\phantom{\hspace{2em}}2.0~cm\phantom{\hspace{2em}}\phantom{\hspace{2em}}}} & \multicolumn{3}{c}{\underline{\phantom{\hspace{2em}}\phantom{\hspace{2em}}1.3~cm\phantom{\hspace{2em}}\phantom{\hspace{2em}}}} & \multicolumn{3}{c}{\underline{\phantom{\hspace{2em}}\phantom{\hspace{2em}}0.7~cm\phantom{\hspace{2em}}\phantom{\hspace{2em}}}} \\
\colhead{Array} & \colhead{Number} & \colhead{$\langle S_\nu\rangle$} & \colhead{$\sigma$} & \colhead{Number} & \colhead{$\langle S_\nu\rangle$} & \colhead{$\sigma$} & \colhead{Number} & \colhead{$\langle S_\nu\rangle$} & \colhead{$\sigma$} \\
\colhead{} & \colhead{of Epochs} & \colhead{(Jy)} & \colhead{(Jy)} & \colhead{of Epochs} & \colhead{(Jy)} & \colhead{(Jy)} & \colhead{of Epochs} & \colhead{(Jy)} & \colhead{(Jy)} 
}
\startdata
A       & 33 &  0.86 &  0.14 & 32 &  0.96 &  0.17 & 29 &  1.03 &  0.24 \\ 
BnA     &  6 &  0.77 &  0.24 &  6 &  0.83 &  0.23 &  7 &  0.83 &  0.35 \\ 
B       & 26 &  0.81 &  0.11 & 26 &  0.91 &  0.12 & 26 &  0.97 &  0.13 \\ 
CnB     &  7 &  0.73 &  0.09 &  7 &  0.83 &  0.17 &  7 &  0.95 &  0.38 \\ 
C       & 21 &  0.85 &  0.12 & 20 &  0.97 &  0.14 & 20 &  1.13 &  0.15 \\ 
DnC     &  5 &  0.94 &  0.11 &  5 &  0.94 &  0.18 &  5 &  1.01 &  0.21 \\ 
D       & 17 &  0.85 &  0.13 & 28 &  0.93 &  0.17 & 27 &  1.00 &  0.16 \\ 
\enddata
\tablecomments{The standard deviation is calculated from the spread of measured flux densities for that array.  BnA refers to the hybrid array of the VLA
with the northern arm in the A configuration (see \S \ref{res}).}
\end{deluxetable}

%%%%%%%%%%%%%%%%%%%%%  END  TABLE 4 %%%%%%%%%%%%%%%%%%%%%%%

\end{document}

%% file: Herrnstein.tab2.dat
20000621 & 0 & ... & 0.85\p0.05 & 0.87\p0.09 & ...~~~~~~  \\
20000629 & 8 & ... & 0.90\p0.06 & 0.84\p0.07 & ...~~~~~~  \\
20000707 & 16 & ... & 0.94\p0.06 & 0.83\p0.08 & ...~~~~~~  \\
20000717 & 26 & ... & 1.21\p0.08 & 0.13\p0.14 & ...~~~~~~  \\
20000720 & 29 & ... & 1.00\p0.08 & 0.87\p0.10 & ...~~~~~~  \\
20000806 & 47 & ... & 0.72\p0.05 & 0.04\p0.09 & ...~~~~~~  \\
20000814 & 54 & ... & 0.64\p0.06 & 0.99\p0.09 & ...~~~~~~  \\
20000908 & 80 & ... & 0.86\p0.04 & 0.04\p0.08 & ...~~~~~~  \\
20000913 & 85 & ... & 0.60\p0.06 & ... & ...~~~~~~  \\
20000926 & 98 & ... & 0.67\p0.05 & 0.93\p0.09 & ...~~~~~~  \\
20001005 & 107 & 0.86\p0.12 & 0.90\p0.07 & 0.89\p0.09 & 0.01\p0.15  \\
20001009 & 111 & 0.90\p0.06 & 0.98\p0.06 & 1.13\p0.09 & 0.22\p0.10  \\
20001017 & 119 & ... & 0.96\p0.06 & 1.02\p0.08 & ...~~~~~~  \\
20001024 & 126 & 0.88\p0.06 & 0.94\p0.08 & 0.94\p0.13 & 0.08\p0.14  \\
20001031 & 133 & 0.97\p0.06 & 1.00\p0.07 & 1.08\p0.14 & 0.10\p0.13  \\
20001105 & 138 & 1.15\p0.07 & 1.26\p0.07 & 1.34\p0.16 & 0.17\p0.12  \\
20001110 & 143 & 0.89\p0.05 & 0.99\p0.06 & 1.24\p0.16 & 0.29\p0.13  \\
20001116 & 149 & 0.93\p0.06 & 1.06\p0.09 & 1.23\p0.20 & 0.28\p0.15  \\
20001118 & 151 & 0.88\p0.06 & ... & ... & ...~~~~~~  \\
20001127 & 160 & 0.89\p0.06 & 0.86\p0.06 & 0.84\p0.07 & --0.05\p0.10  \\
20001203 & 166 & 0.83\p0.05 & 0.93\p0.06 & 1.07\p0.10 & 0.24\p0.10  \\
20001209 & 172 & 0.92\p0.05 & 1.00\p0.05 & 1.16\p0.08 & 0.22\p0.08  \\
20001217 & 180 & 1.09\p0.06 & 1.14\p0.07 & 1.07\p0.08 & --0.02\p0.09  \\
20001222 & 185 & 1.05\p0.06 & 1.13\p0.06 & 0.98\p0.09 & --0.04\p0.10  \\
20001229 & 192 & 0.98\p0.05 & 1.03\p0.06 & 0.93\p0.06 & --0.06\p0.08  \\
20010111 & 205 & 0.99\p0.06 & 1.13\p0.06 & 1.08\p0.09 & 0.09\p0.10  \\
20010119 & 213 & 0.89\p0.05 & 0.93\p0.06 & 0.83\p0.10 & --0.03\p0.12  \\
20010126 & 220 & 1.03\p0.07 & 1.08\p0.06 & ... & ...~~~~~~  \\
20010129 & 223 & 0.94\p0.06 & 0.94\p0.07 & ... & ...~~~~~~  \\
20010203 & 228 & 0.82\p0.05 & 0.88\p0.05 & 0.92\p0.08 & 0.11\p0.10  \\
20010214 & 239 & 0.68\p0.04 & 0.69\p0.04 & 0.63\p0.06 & --0.07\p0.10  \\
20010219 & 244 & ... & ... & 0.83\p0.08 & ...~~~~~~  \\
20010221 & 246 & 0.69\p0.04 & 0.79\p0.05 & 0.88\p0.07 & 0.23\p0.09  \\
20010304 & 257 & 0.93\p0.06 & 1.07\p0.06 & 1.02\p0.10 & 0.12\p0.11  \\
20010305 & 258 & ... & ... & 1.09\p0.10 & ...~~~~~~  \\
20010317 & 270 & 1.21\p0.08 & 1.33\p0.13 & ... & ...~~~~~~  \\
20010322 & 275 & 0.91\p0.05 & 1.05\p0.06 & 1.09\p0.14 & 0.22\p0.12  \\
20010401 & 285 & 0.99\p0.06 & 1.04\p0.06 & 1.07\p0.10 & 0.08\p0.10  \\
20010407 & 291 & 0.94\p0.06 & 1.01\p0.06 & 1.17\p0.09 & 0.20\p0.09  \\
20010416 & 300 & 0.77\p0.05 & 0.85\p0.05 & 0.93\p0.09 & 0.18\p0.11  \\
20010422 & 306 & 0.76\p0.04 & 0.88\p0.05 & 0.98\p0.06 & 0.23\p0.07  \\
20010429 & 313 & 0.82\p0.05 & 0.93\p0.06 & 0.96\p0.09 & 0.17\p0.10  \\
20010505 & 319 & 0.79\p0.05 & 0.89\p0.06 & 0.97\p0.08 & 0.20\p0.10  \\
20010511 & 325 & 0.67\p0.05 & 0.79\p0.06 & 0.98\p0.11 & 0.36\p0.12  \\
20010520 & 334 & 0.86\p0.05 & 0.90\p0.06 & 0.93\p0.11 & 0.08\p0.12  \\
20010524 & 338 & 0.75\p0.04 & 0.82\p0.05 & 0.89\p0.08 & 0.16\p0.10  \\
20010528 & 342 & 0.77\p0.05 & 0.81\p0.05 & 0.87\p0.07 & 0.11\p0.10  \\
20010603 & 348 & 0.85\p0.05 & 0.98\p0.05 & 1.05\p0.08 & 0.20\p0.09  \\
20010611 & 356 & 0.69\p0.04 & 0.76\p0.05 & 0.80\p0.09 & 0.15\p0.12  \\
20010617 & 362 & 0.63\p0.04 & 0.67\p0.04 & 0.71\p0.06 & 0.13\p0.10  \\
20010624 & 369 & 0.71\p0.04 & 0.79\p0.05 & 0.90\p0.09 & 0.23\p0.11  \\
20010702 & 377 & 0.69\p0.04 & 0.72\p0.05 & 0.89\p0.10 & 0.22\p0.12  \\
20010709 & 384 & 0.79\p0.05 & 0.84\p0.07 & 1.05\p0.14 & 0.25\p0.13  \\
20010714 & 389 & 0.84\p0.05 & 0.87\p0.07 & 1.03\p0.15 & 0.17\p0.14  \\
20010724 & 399 & 0.89\p0.05 & 0.95\p0.06 & 0.95\p0.09 & 0.06\p0.10  \\
20010805 & 412 & 0.83\p0.05 & 1.03\p0.04 & 1.28\p0.09 & 0.40\p0.09  \\
20010813 & 419 & 0.91\p0.07 & ... & ... & ...~~~~~~  \\
20010824 & 430 & 0.91\p0.05 & 0.97\p0.06 & 0.92\p0.09 & 0.03\p0.11  \\
20010831 & 437 & 1.10\p0.06 & 1.21\p0.06 & 1.26\p0.10 & 0.14\p0.09  \\
20010901 & 438 & 1.05\p0.06 & 1.23\p0.08 & 1.34\p0.10 & 0.22\p0.09  \\
20010909 & 446 & 0.92\p0.05 & 0.90\p0.05 & 0.89\p0.11 & --0.05\p0.12  \\
20010926 & 463 & 0.89\p0.05 & 0.80\p0.04 & 0.87\p0.07 & --0.04\p0.09  \\
20010928 & 465 & 0.88\p0.05 & 0.86\p0.05 & 0.92\p0.07 & 0.04\p0.09  \\
20011003 & 470 & 1.00\p0.06 & 1.00\p0.06 & 1.06\p0.12 & 0.05\p0.12  \\
20011011 & 479 & 0.91\p0.05 & 1.16\p0.07 & 1.10\p0.10 & 0.20\p0.10  \\
20011018 & 486 & 0.87\p0.05 & 1.10\p0.06 & 1.21\p0.09 & 0.31\p0.09  \\
20011025 & 493 & 0.64\p0.04 & 0.82\p0.05 & 0.86\p0.08 & 0.29\p0.10  \\
20011101 & 500 & 0.74\p0.05 & 0.84\p0.05 & 0.95\p0.10 & 0.24\p0.11  \\
20011109 & 508 & 0.91\p0.06 & 1.04\p0.06 & 1.18\p0.10 & 0.25\p0.10  \\
20011118 & 515 & 0.88\p0.06 & 1.04\p0.07 & 1.17\p0.09 & 0.27\p0.09  \\
20011201 & 530 & 0.93\p0.05 & 1.08\p0.07 & 0.94\p0.07 & 0.01\p0.09  \\
20011208 & 537 & 1.24\p0.08 & 1.19\p0.10 & 1.48\p0.12 & 0.17\p0.10  \\
20011217 & 546 & 0.96\p0.06 & 1.18\p0.07 & 0.93\p0.06 & --0.05\p0.08  \\
20011223 & 552 & 1.07\p0.06 & 1.07\p0.07 & 1.25\p0.09 & 0.15\p0.09  \\
20020112 & 572 & 0.85\p0.06 & 0.87\p0.07 & 0.96\p0.08 & 0.11\p0.10  \\
20020119 & 579 & 0.92\p0.05 & 0.98\p0.05 & 1.02\p0.10 & 0.11\p0.10  \\
20020203 & 594 & 0.93\p0.05 & 1.07\p0.06 & 1.14\p0.11 & 0.21\p0.10  \\
20020211 & 602 & 1.02\p0.07 & 1.18\p0.07 & 1.29\p0.14 & 0.23\p0.11  \\
20020218 & 609 & 0.89\p0.06 & 1.02\p0.06 & ... & ...~~~~~~  \\
20020226 & 617 & 0.62\p0.05 & 0.72\p0.06 & 0.95\p0.14 & 0.39\p0.15  \\
20020304 & 623 & 0.90\p0.05 & 1.01\p0.05 & 1.06\p0.07 & 0.16\p0.08  \\
20020311 & 630 & 0.74\p0.04 & 0.88\p0.05 & 0.96\p0.07 & 0.25\p0.09  \\
20020318 & 637 & 0.72\p0.04 & 0.80\p0.05 & 0.87\p0.09 & 0.19\p0.11  \\
20020325 & 644 & 0.74\p0.04 & 0.84\p0.05 & 0.91\p0.09 & 0.20\p0.11  \\
20020404 & 654 & 0.72\p0.04 & 0.79\p0.05 & 0.93\p0.08 & 0.23\p0.10  \\
20020411 & 661 & 0.82\p0.05 & 0.91\p0.05 & 1.18\p0.10 & 0.34\p0.10  \\
20020417 & 667 & 0.74\p0.05 & 0.84\p0.06 & 1.05\p0.07 & 0.34\p0.09  \\
20020424 & 674 & 0.71\p0.04 & 0.86\p0.06 & 1.00\p0.07 & 0.32\p0.09  \\
20020505 & 685 & 0.68\p0.04 & 0.82\p0.05 & 0.91\p0.07 & 0.28\p0.10  \\
20020520 & 700 & 0.75\p0.04 & 0.81\p0.05 & 0.82\p0.09 & 0.10\p0.11  \\
20020526 & 706 & 0.78\p0.05 & 0.94\p0.05 & 0.89\p0.09 & 0.15\p0.11  \\
20020602 & 713 & 0.78\p0.05 & 0.85\p0.05 & 0.84\p0.08 & 0.08\p0.11  \\
20020607 & 718 & 0.66\p0.04 & 0.75\p0.05 & 0.73\p0.07 & 0.09\p0.11  \\
20020616 & 727 & 0.77\p0.05 & 0.87\p0.05 & 0.89\p0.09 & 0.15\p0.11  \\
20020623 & 734 & 0.92\p0.05 & 1.12\p0.06 & 1.26\p0.10 & 0.30\p0.09  \\
20020701 & 742 & 0.86\p0.05 & 0.93\p0.06 & 0.87\p0.06 & 0.01\p0.09  \\
20020706 & 747 & 0.75\p0.05 & 0.84\p0.05 & 0.93\p0.08 & 0.20\p0.10  \\
20020725 & 766 & 0.82\p0.05 & 0.91\p0.05 & 0.80\p0.07 & --0.01\p0.10  \\
20020801 & 773 & 0.91\p0.06 & 1.00\p0.06 & 1.18\p0.09 & 0.24\p0.09  \\
20020808 & 780 & 0.76\p0.05 & 0.85\p0.06 & 0.98\p0.08 & 0.24\p0.10  \\
20020815 & 787 & 0.74\p0.05 & 0.85\p0.05 & 0.96\p0.08 & 0.24\p0.10  \\
20020820 & 793 & 0.82\p0.05 & 0.94\p0.06 & 1.12\p0.08 & 0.29\p0.09  \\
20020901 & 805 & 0.85\p0.06 & 1.01\p0.07 & 1.25\p0.09 & 0.36\p0.09  \\
20020910 & 813 & 0.86\p0.06 & 1.02\p0.07 & 1.25\p0.10 & 0.35\p0.09  \\
20020918 & 821 & 0.76\p0.05 & 0.89\p0.06 & 1.12\p0.09 & 0.36\p0.10  \\
20020928 & 832 & 0.73\p0.05 & 0.86\p0.05 & 1.06\p0.09 & 0.35\p0.10  \\
20021003 & 837 & 0.88\p0.05 & 1.18\p0.06 & 1.86\p0.18 & 0.70\p0.10  \\
20021012 & 846 & 0.79\p0.05 & 0.90\p0.05 & 1.09\p0.08 & 0.31\p0.09  \\
20021018 & 852 & 0.69\p0.04 & 0.82\p0.05 & 1.08\p0.11 & 0.42\p0.11  \\
20021024 & 858 & 0.75\p0.05 & 0.95\p0.05 & 1.28\p0.11 & 0.51\p0.10  \\
20021103 & 868 & 0.75\p0.05 & 0.94\p0.06 & 1.32\p0.16 & 0.53\p0.13  \\
20021118 & 883 & 0.79\p0.05 & 0.98\p0.06 & 1.17\p0.11 & 0.37\p0.10  \\
20021129 & 894 & 0.91\p0.05 & 1.00\p0.06 & 1.12\p0.09 & 0.19\p0.10  \\
20021205 & 900 & 0.97\p0.06 & 1.13\p0.06 & 1.31\p0.11 & 0.28\p0.09  \\
20021213 & 908 & 0.85\p0.05 & 1.02\p0.06 & 1.22\p0.10 & 0.34\p0.10  \\
20021219 & 914 & 0.83\p0.05 & 1.05\p0.07 & 1.23\p0.11 & 0.38\p0.10  \\
20021228 & 923 & 1.08\p0.06 & 1.25\p0.07 & 1.07\p0.10 & 0.02\p0.10  \\
20030104 & 930 & 0.94\p0.05 & 1.01\p0.06 & 1.29\p0.09 & 0.30\p0.09  \\
20030113 & 939 & 1.13\p0.06 & 1.22\p0.07 & 1.40\p0.11 & 0.20\p0.09  \\
20030214 & 971 & 0.87\p0.05 & 1.23\p0.07 & 1.16\p0.10 & 0.30\p0.10  \\
20030317 & 1002 & 0.79\p0.04 & 1.13\p0.06 & 0.93\p0.08 & 0.21\p0.10  \\
20030416 & 1032 & 0.80\p0.05 & 1.08\p0.06 & 1.00\p0.09 & 0.27\p0.10  \\
20030519 & 1065 & 0.86\p0.05 & 1.08\p0.06 & 1.22\p0.10 & 0.34\p0.10  \\
20030616 & 1094 & 0.88\p0.09 & 1.06\p0.12 & 1.70\p0.20 & 0.62\p0.14  \\
20030716 & 1124 & 1.25\p0.08 & 1.64\p0.10 & 1.87\p0.16 & 0.39\p0.10  \\
20030816 & 1155 & 0.98\p0.06 & 1.13\p0.07 & 1.34\p0.11 & 0.30\p0.10  \\
20030914 & 1184 & 0.93\p0.06 & 1.04\p0.07 & 1.23\p0.12 & 0.26\p0.11  \\
20031014 & 1214 & 1.32\p0.08 & 1.36\p0.09 & 1.73\p0.15 & 0.25\p0.10  \\

%% file: Herrnstein.bbl
\begin{thebibliography}{27}
\expandafter\ifx\csname natexlab\endcsname\relax\def\natexlab#1{#1}\fi

\bibitem[{{Baganoff}(2003)}]{bag03}
{Baganoff}, F.~K. 2003, AAS/High Energy Astrophysics Division, 35

\bibitem[{{Baganoff} {et~al.}(2001)}]{bag01}
{Baganoff}, F.~K. {et~al.} 2001, \nat, 413, 45

\bibitem[{{Bower} {et~al.}(2004){Bower}, {Falcke}, {Herrnstein}, {Zhao},
  {Goss}, \& {Backer}}]{bower04}
{Bower}, G.~C., {Falcke}, H., {Herrnstein}, R.~M., {Zhao}, J.-H., {Goss},
  W.~M., \& {Backer}, D.~C. 2004, submitted to Science

\bibitem[{{Bower} {et~al.}(2002){Bower}, {Falcke}, {Sault}, \&
  {Backer}}]{bower02}
{Bower}, G.~C., {Falcke}, H., {Sault}, R.~J., \& {Backer}, D.~C. 2002, \apj,
  571, 843

\bibitem[{{Brown} \& {Lo}(1982)}]{bro82}
{Brown}, R.~L. \& {Lo}, K.~Y. 1982, \apj, 253, 108

\bibitem[{{Falcke}(1999)}]{fal99}
{Falcke}, H. 1999, in ASP Conf. Ser. 186: The Central Parsecs of the Galaxy,
  113

\bibitem[{{Falcke} {et~al.}(1998){Falcke}, {Goss}, {Matsuo}, {Teuben}, {Zhao},
  \& {Zylka}}]{fal98}
{Falcke}, H., {Goss}, W.~M., {Matsuo}, H., {Teuben}, P., {Zhao}, J., \&
  {Zylka}, R. 1998, \apj, 499, 731

\bibitem[{{Genzel} {et~al.}(2003){Genzel}, {Sch\"{o}del}, {Ott}, {Eckart},
  {Alexander}, {Lacombe}, \& {Aschenbach}}]{gen03}
{Genzel}, R., {Sch\"{o}del}, R., {Ott}, T., {Eckart}, A., {Alexander}, T.,
  {Lacombe}, F.~D.~R., \& {Aschenbach}, B. 2003, \nat, in press

\bibitem[{{Ghez} {et~al.}(2003{\natexlab{a}}){Ghez}, {Becklin}, {Duch\^{e}ne},
  {Hornstein}, {Morris}, {Salim}, \& {Tanner}}]{ghe03}
{Ghez}, A.~M., {Becklin}, E., {Duch\^{e}ne}, G., {Hornstein}, S., {Morris}, M.,
  {Salim}, S., \& {Tanner}, A. 2003{\natexlab{a}}, in Astron. Nachr., Vol. 324,
  No. S1, Special Supplement ``The central 300 parsecs of the Milky Way'', Eds.
  A. Cotera, H. Falcke, T. R. Geballe, S. Markoff,, Vol. 324, 527--533

\bibitem[{{Ghez} {et~al.}(2003{\natexlab{b}})}]{ghe03b}
{Ghez}, A.~M. {et~al.} 2003{\natexlab{b}}, \apjl, 586, L127

\bibitem[{{Goldwurm} {et~al.}(2003){Goldwurm}, {Brion}, {Goldoni}, {Ferrando},
  {Daigne}, {Decourchelle}, {Warwick}, \& {Predehl}}]{goldwurm03}
{Goldwurm}, A., {Brion}, E., {Goldoni}, P., {Ferrando}, P., {Daigne}, F.,
  {Decourchelle}, A., {Warwick}, R.~S., \& {Predehl}, P. 2003, \apj, 584, 751

\bibitem[{{Loeb}(2004)}]{loeb04}
{Loeb}, A. 2004, \mnras, in press (astro-ph/0311512)

\bibitem[{{Marscher} \& {Gear}(1985)}]{marscher85}
{Marscher}, A.~P. \& {Gear}, W.~K. 1985, \apj, 298, 114

\bibitem[{{Menten} {et~al.}(1997){Menten}, {Reid}, {Eckart}, \&
  {Genzel}}]{men97}
{Menten}, K.~M., {Reid}, M.~J., {Eckart}, A., \& {Genzel}, R. 1997, \apjl, 475,
  L111

\bibitem[{{Porquet} {et~al.}(2003){Porquet}, {Predehl}, {Aschenbach}, {Grosso},
  {Goldwurm}, {Goldoni}, {Warwick}, \& {Decourchelle}}]{porquet03}
{Porquet}, D., {Predehl}, P., {Aschenbach}, B., {Grosso}, N., {Goldwurm}, A.,
  {Goldoni}, P., {Warwick}, R.~S., \& {Decourchelle}, A. 2003, \aap, 407, L17

\bibitem[{{Press} {et~al.}(1989){Press}, {Flannery}, {Teukolsky}, \&
  {Vetterling}}]{pre89}
{Press}, W.~H., {Flannery}, B.~P., {Teukolsky}, S.~A., \& {Vetterling}, W.~T.
  1989, {Numerical recipes in C. The art of scientific computing} (Cambridge:
  University Press, 1989)

\bibitem[{{Rickett}(1990)}]{rickett90}
{Rickett}, B.~J. 1990, \araa, 28, 561

\bibitem[{{Sch{\" o}del} {et~al.}(2002)}]{sch02}
{Sch{\" o}del}, R. {et~al.} 2002, \nat, 419, 694

\bibitem[{{Serabyn} {et~al.}(1997){Serabyn}, {Carlstrom}, {Lay}, {Lis},
  {Hunter}, \& {Lacy}}]{ser97}
{Serabyn}, E., {Carlstrom}, J., {Lay}, O., {Lis}, D.~C., {Hunter}, T.~R., \&
  {Lacy}, J.~H. 1997, \apjl, 490, L77

\bibitem[{{Tsuboi} {et~al.}(1999){Tsuboi}, {Miyazaki}, \& {Tsutsumi}}]{tsu99}
{Tsuboi}, M., {Miyazaki}, A., \& {Tsutsumi}, T. 1999, in ASP Conf. Ser. 186:
  The Central Parsecs of the Galaxy, 105

\bibitem[{{Tsutsumi} {et~al.}(2002){Tsutsumi}, {Miyazaki}, \& {Tsuboi}}]{tsu02}
{Tsutsumi}, T., {Miyazaki}, A., \& {Tsuboi}, M. 2002, in American Astronomical
  Society Meeting 200, \#44.09

\bibitem[{{Wright} \& {Backer}(1993)}]{wri93}
{Wright}, M.~C.~H. \& {Backer}, D.~C. 1993, \apj, 417, 560

\bibitem[{{Zhao} {et~al.}(2001){Zhao}, {Bower}, \& {Goss}}]{zha01}
{Zhao}, J.-H., {Bower}, G.~C., \& {Goss}, W.~M. 2001, \apjl, 547, L29

\bibitem[{{Zhao} {et~al.}(1992){Zhao}, {Goss}, {Lo}, \& {Ekers}}]{zha92}
{Zhao}, J.-H., {Goss}, W.~M., {Lo}, K.-Y., \& {Ekers}, R.~D. 1992, in ASP Conf.
  Ser. 31: Relationships Between Active Galactic Nuclei and Starburst Galaxies,
  295

\bibitem[{{Zhao} {et~al.}(2004){Zhao}, {Herrnstein}, {Bower}, {Goss}, \&
  {Liu}}]{zha04}
{Zhao}, J.-H., {Herrnstein}, R.~M., {Bower}, G.~C., {Goss}, W.~M., \& {Liu},
  S.~M. 2004, \apjl, in press (astro-ph/0401508)

\bibitem[{{Zhao} {et~al.}(2003){Zhao}, {Young}, {Herrnstein}, {Ho}, {Tsutsumi},
  {Lo}, {Goss}, \& {Bower}}]{zha03}
{Zhao}, J.-H., {Young}, K.~H., {Herrnstein}, R.~M., {Ho}, P.~T.~P., {Tsutsumi},
  T., {Lo}, K.~Y., {Goss}, W.~M., \& {Bower}, G.~C. 2003, \apjl, 586, L29

\bibitem[{{Zylka} {et~al.}(1992){Zylka}, {Mezger}, \& {Lesch}}]{zyl92}
{Zylka}, R., {Mezger}, P.~G., \& {Lesch}, H. 1992, \aap, 261, 119

\end{thebibliography}
